\documentclass[sigconf]{acmart}
\usepackage{enumitem}
\usepackage{geometry}
\usepackage{graphicx}
\usepackage{float}

\AtBeginDocument{%
  \providecommand\BibTeX{{%
    \normalfont B\kern-0.5em{\scshape i\kern-0.25em b}\kern-0.8em\TeX}}}

\copyrightyear{}
\acmYear{}
\acmDOI{}

\acmBooktitle{ } 
\acmPrice{}
\acmISBN{}

\begin{document}

\title{PADME-SoSci: A Platform for Analytics and Distributed Machine Learning for the Social Sciences}


\author{Zeyd~Boukhers$^{1,2}$ \quad Arnim~Bleier$^{3}$ \quad Yeliz~Ucer~Yediel$^{1,5}$ \quad Mio Hienstorfer-Heitmann$^{3}$ \quad Mehrshad~Jaberansary$^{2,4}$ \quad Adamantios~Koumpis$^{2,4}$ \quad Oya~Beyan$^{1,2,4}$}
\affiliation{%
  {$^1$Fraunhofer Institute for Applied Information Technology -FIT-, Sankt Augustin, Germany}\\
  {$^2$University Hospital Cologne, Cologne, Germany}\\
  {$^3$GESIS – Leibniz Institute for the Social Sciences, Cologne, Germany}\\
  {$^4$University of Cologne, Cologne, Germany}\\
  {$^5$RWTH Aachen University, Aachen, Germany}\\
  \country{Germany}
}
\email{{zeyd.boukhers, yeliz.ucer.yediel} @fit.fraunhofer.de}
\email{{arnim.bleier, mio.hienstorfer-heitmann}@gesis.org}
\email{{mehrshad.jaberansary, adamantios.koumpis, oya.beyan}@uk-koeln.de}

\renewcommand{\shortauthors}{Boukhers et al.}

\begin{abstract}

Data privacy and ownership are significant in social data science, raising legal and ethical concerns. Sharing and analyzing data is difficult when different parties own different parts of it. An approach to this challenge is to apply de-identification or anonymization techniques to the data before collecting it for analysis. However, this can reduce data utility and increase the risk of re-identification. To address these limitations, we present \textbf{PADME}, a distributed analytics tool that federates model implementation and training. \textbf{PADME} uses a federated approach where the model is implemented and deployed by all parties and visits each data location incrementally for training. This enables the analysis of data across locations while still allowing the model to be trained as if all data were in a single location. Training the model on data in its original location preserves data ownership. Furthermore, the results are not provided until the analysis is completed on all data locations to ensure privacy and avoid bias in the results.

\end{abstract}



\maketitle

\section{Introduction}
\label{sec:intro}

Data privacy and ownership are crucial in social data science as the data often includes sensitive personal information. For example, it is expected that political survey data will typically be gathered by various parties across different groups, but sharing may not occur due to privacy considerations. As a result, each party independently analyzes their own data and only shares the outcome. This approach can be limited as the aggregated outcome may not accurately reflect the entire data. An alternative solution is to mask personal information, but this can also have limitations as masked attributes may be important for analysis.

To address this issue, we present \textbf{PADME}\footnote{\url{https://padme-analytics.de/}}~\cite{welten2022privacy}, a distributed analytics tool that enables the training of models on large-scale datasets without the need to centralize the data. \textbf{PADME} trains the analytic model incrementally by visiting each data station one by one. This ensures that the data remains decentralized and protected from unauthorized access. Additionally, \textbf{PADME} only produces output after visiting all data stations, thereby avoiding the sharing of derived knowledge that could potentially reveal information about the data within a specific station. This provides an added layer of security and ensures that the data remains confidential. To further guarantee the security and privacy of the data, it is recommended that the model is implemented in a federated manner. This ensures that the data is protected by multiple parties and that any potential vulnerabilities are identified and addressed.
\section{PADME-SoSci}



\begin{figure}[ht!]
    \centering
    \includegraphics[width=\linewidth]{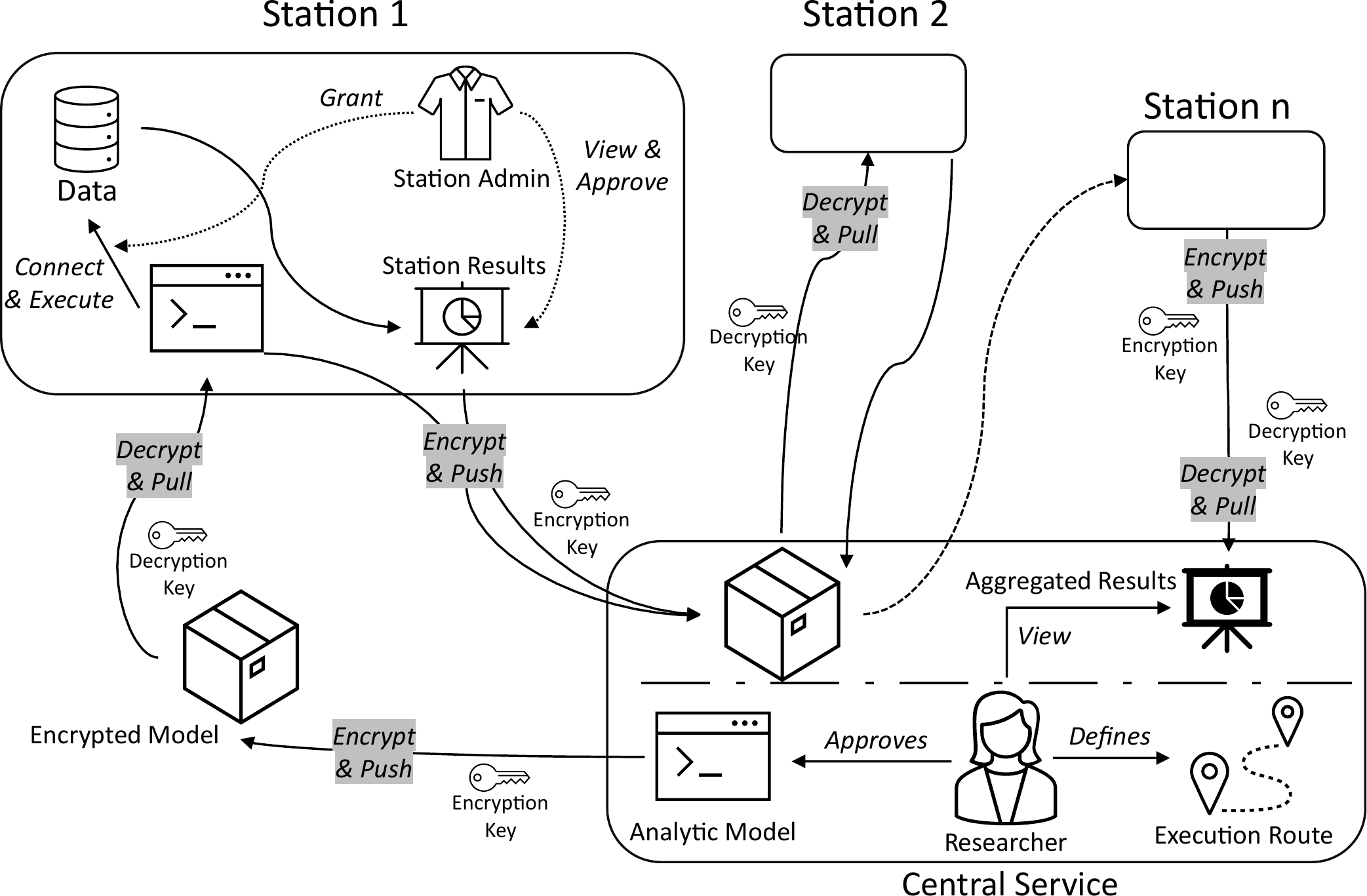}
    \caption{\centering An overview of the distributed analytic tool with $n$ data stations}
    \label{fig:overview}
\end{figure}

The distributed analytics tool depicted in Figure~\ref{fig:overview} has two primary components: the \emph{Service Center} and the \emph{Data Stations}. The \emph{Service Center} acts as the central hub for the entire analytics process. It oversees the development and approval of analytic models, manages the authentication and authorization of scientists and researchers, and ensures the proper execution of analytics tasks. The process of developing and approving the analytic models involves the collaboration and agreement of both the researcher and the data owners. This consensus mechanism ensures that the parameters and architecture of the global model are agreed upon by all parties. The \emph{Service Center} also ensures the security of the analytics process by securely packaging the analytics model into containers and managing the execution routes through authentication. This minimizes the risk of unauthorized access or misuse of the analytic models. Additionally, the \emph{Service Center} employs encryption techniques to protect the privacy and ownership of data throughout the entire process. The encrypted containers carrying the analytics models are then transferred to the next target station.


The \emph{Data Station} operates independently of the \emph{Service Center}, but it remains closely connected to it through a secure communication channel. The station pulls the encrypted container from the \emph{Service Center} after authenticating itself. The station admin then provides the necessary connection information to allow the decrypted container to execute the analytics model. The station's unique key is a crucial security feature that ensures that only the designated station can execute the analytic model within the decrypted container. After the analytics task is completed, the Station Admin reviews the results and, upon approval, packages them into a secure container using the station's specific key. This process is repeated for each station in the route until the final results are available in an encrypted container. The authenticated scientist can then retrieve, decrypt and view the aggregated results.

\subsection{Prerequisites}
To effectively utilize the distributed analytic tool, compliance with the following prerequisites is crucial:

\begin{itemize}[leftmargin=*]
    \item \textbf{Data standardization:} All used data must be in a standardized format that the model is designed to work with.

    \item \textbf{Distributed model:} The analytic model must be capable of being distributed. For example, the Latent Dirichlet Allocation (LDA) model requires the entire vocabulary of the corpus before running, making it unsuitable for direct use with PADME. 
    
    \item \textbf{Computational resources:} Every data station needs to have adequate computational resources to execute the analytic model. 

\end{itemize}



\section{Use Cases}

\subsection{Sentiment Analysis}


Online reproducibility services, such as mybinder.org, have become popular in the Computational Social Science community \cite{nbs_2022}.  These services allow researchers to share complex computational analysis pipelines in the form of notebooks that can be easily executed in a browser without the need for additional software installation. However, the pipelines are currently limited to public data. In this paper, we demonstrate the use of PADME for a sentiment analysis test within the following prototypical workflow\footnote{Originally discussed on GitHub \url{https://github.com/gesiscss/btw17_sample_scripts/issues/4}}:

\begin{enumerate}[leftmargin=*]
    \item Schema data is created and publicly shared for a sensitive dataset that can not be shared.
    \item An interested researcher that would like to access the sensitive dataset develops an analysis using the publicly available schema data.
    \item The researcher submits the analysis to PADME to execute the analysis of the private data. 
    \item An exit control on the analysis results is performed, and the results of the analysis are sent back to the researcher if the exit control is passed.
\end{enumerate}
The publicly available part of the German Federal Election 2017 Twitter dataset\cite{stier2018systematically} (DBK: ZA6926) is used for this use case.

\subsection{Supervised Author Name Disambiguation}
In this demonstration, we are showcasing the capability of PADME by distributing the Author Name Disambiguation (AND) task to be carried out at two separate stations, each of which holds a portion of the data. The data used for AND is typically open, but the purpose of this demonstration is to highlight the versatility of PADME in handling various types of data and analytical models. In particular, we are utilizing a supervised neural network model~\cite{boukhers2022whois} that has been trained on a preprocessed DBLP dataset\footnote{\url{https://doi.org/10.5281/zenodo.7506562}}. The DBLP dataset has been split and distributed between the two stations for this demonstration.

\section*{Acknowledgement}
This work was partially supported by the project NFDI4DS\footnote{\url{https://gepris.dfg.de/gepris/projekt/460234259}} with grant number \emph{460234259}.

\bibliographystyle{ACM-Reference-Format}
\bibliography{sample-base}

\end{document}